# Machine Learning based Intelligent Cognitive Network using Fog Computing

Jingyang Lu*[a], Lun Li[a], Genshe Chen[a], Dan Shen[a], Khanh Pham[b], Erik Blasch[c]

[a]Intelligent Fusion Technology, Inc., 20271 Goldenrod Ln, Germantown, MD, 20876;
[b]Air Force Research Lab, Kirtland AFB, NM, 87117;
[c]Air Force Research Lab, Rome, NY, 13441.

## ABSTRACT

In this paper, a Cognitive Radio Network (CRN) based on artificial intelligence is proposed to distribute the limited radio spectrum resources more efficiently. The CRN framework can analyze the time-sensitive signal data close to the signal source using fog computing with different types of machine learning techniques. Depending on the computational capabilities of the fog nodes, different features and machine learning techniques are chosen to optimize spectrum allocation. Also, the computing nodes send the periodic signal summary which is much smaller than the original signal to the cloud so that the overall system spectrum source allocation strategies are dynamically updated. Applying fog computing, the system is more adaptive to the local environment and robust to spectrum changes. As most of the signal data is processed at the fog level, it further strengthens the system security by reducing the communication burden of the communications network.

## 1. INTRODUCTION

With large growth of wireless services and applications, there exists an exponential demand for the available radio spectrum. Some radio spectrum resources are sold to the primary users, which sometimes are not fully used. In some countries, most of available spectrum resources have been fully utilized resulting in the spectrum scarcity problem. On the other hand, a large amount of the licensed spectrum experiences low utilization based on the recent studies on the actual spectrum utilization measurements, which indicates inefficiency issues existing in the radio spectrum resources allocation. In the dynamic spectrum access (DSA) system, the opportunistic unlicensed secondary users sense the channel to find the available radio spectrum over which they can transmit signal without interfering with the primary users. The secondary users of such capability utilize an intelligent method called Cognitive Radio (CR). There have been different cognitive radio applications developed such as spectrum sensing[1], autonomous learning, channel estimation and data detection[2], user cooperation, modeling, target tracking [3][4][5][6] and reasoning.

The inefficiency of spectrum use has triggered exciting activities in engineering, economic, and regulation communities to develop improved spectrum management and spectrum sharing. The concept of Cognitive Radio (CR) was first proposed by Mitola and Gerald Q. Maguire[7]. The opportunistic spectrum access model[8] is a landmark example. Given the constraints of collision probability and overlapping time, the scenario in which secondary users can opportunistically access the unused spectrum vacated by idle primary users was investigated. Three spectrum access schemes using different sensing, back-off, and transmission mechanisms are shown to achieve indistinguishable secondary performance. Once detecting one or multiple spectrum holes, the secondary users reconfigure their transmission parameters such as the carrier frequency, modulation scheme, etc. to operate in the detected spectrum holes. Different dynamic spectrum access models such as dynamic exclusive use model, open sharing model, and hierarchical access model are methods to improve spectrum efficiency[9]. The fundamental capacity limits and associated transmission techniques for different wireless network design paradigms are surveyed[10]. Based on the prior rules related to spectrum usage, the interaction between cognitive radios' signals and the transmissions of noncognitive nodes is of interest. Besides the opportunistic spectrum access model, the concurrent spectrum access model is investigated by many researchers, where second users coexist with primary users within a licensed band. Under the constraints that the interference caused by second users' transmitted signal is within the primary users' reasonable threshold, three tasks including radio-scene analysis, channel-state estimation, and transmit-power control and dynamic spectrum management are compared[11].

Regarding spectrum detection and allocation, many different approaches have been utilized such as machine learning, artificial intelligence, model prediction[12],fuzzy logic reasoning[13], greedy search[14], and dynamic programming[15] . As for machine learning, the system first learns the pattern of the object through the training samples, then the system can perform the operation on the data which is not encountered before. Machine learning has wide application areas such as target detection, image processing[16], and pattern recognition[17]. For example, neural networks (NN) can learn the patterns from the structured training data with different features such as power spectrum energy, wave form, modulation, and etc. through supervised or unsupervised methods, which can be seen in Figure 1. Based on the weights achieved from training process, each node or the whole system can predict or conduct the corresponding spectrum decision.

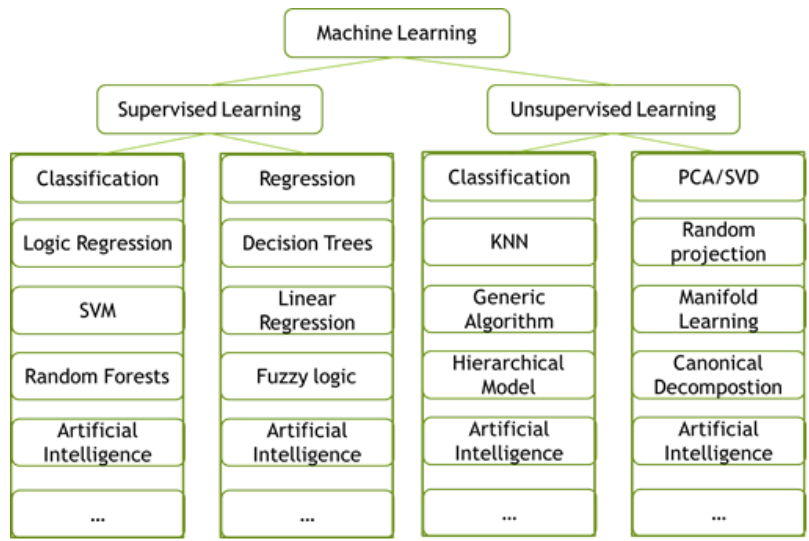

Figure 1 Different Machine Learning Approaches

A contemporary technique is to combine machine learning and game-theoretic methods. Since the over complexity of the selected model may lead to overfitting, there is a need to adaptively learn the situation when other users try to access the system. In the scenario of *opportunistic spectrum access*, it is aimed to find the spectrum holes to transmit the signal without interference the primary users. Based on the non-cooperative independent decisions[18], two dynamic spectrum access algorithms including Hungarian and greedy search are proposed allowing the second users to access the available spectrum. The greedy algorithm can reduce the searching complexity to linear level by selecting the best candidate in each iteration is investigated[19],[20]. It is supposed that the channel availability statistics[21] and the number of secondary users are unknown, where the problem is formulated as a *stochastic game*, which can also be applied in the threat detection and situation awareness[22] [23]. The Nash equilibrium is first achieved based on perfect environment knowledge. A stochastic learning automata-based channel selection algorithm is proposed so that the second users can approach a Nash equilibrium point.

A cognitive network consists of multiple cognitive radio nodes, where different nodes can conduct the spectrum sensing and choose the best spectrum to transmit the signal. Two methods include a centralized approach where all nodes communicate with the cloud or a decentralized approach where only a few fog nodes communicate between the cloud and the other nodes. These two frameworks are illustrated in Figure 2. It is necessary for each cognitive node to communicate with each other, because without information from the other nodes, the node is prone to have a local optimal decision. It will increase the system's spectrum efficiency if all the spectrum sensing information is accessible to the central nodes or cloud. If the central node has the access to all the information, it is capable of achieving the global optimal spectrum allocation strategies. However there also exist some limitations such as communication burden, system security, and system's response to the change of local environment when all fog nodes send the information to the central node or cloud which has more power computation capability. Since each node keeps receiving signal from its nearby neighbors, a large amount of data needs to be transmitted to the cloud even in the case the transmitting signal is compressed before transmitting to cloud[24][25][26]. In some cases, some signals may of no meaning to be sent back to the central node in terms of the spectrum detection or allocation. Also, during the process that the signal is being sent back, the signal may be jammed or compromised by jammers or neutral users either on purpose or not, which will bring interference to the primary users' signal transmission.

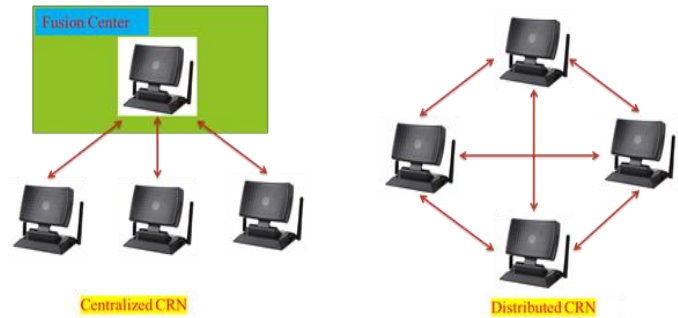

Figure 2 Spectrum Sensing Framework

Cognitive radio methods are promising techniques that can mitigate the spectrum scarcity by enabling the second users to access the limited spectrum resources licensed by the primary users. However, *spectrum sensing security* needs to be considered, as the false information injected into the system can severely interference the spectrum sensing process and reduce the channel availability to the unlicensed legitimate users[27]. The cognitive radio attack surface is studied from the perspective of primary user emulation attacks, which demonstrates that the spectrum efficiency can drop largely by emulating the incumbent signal. The corresponding defending strategy called *localization-based defense* is proposed to defense this sort of attack strategies, which can detect the attack by estimating the incumbent transmitter's location and its signal characteristics. There are also other types of attack strategies which take advantage of the system configuration to inject the false information that the traditional Chi-Square detector cannot detect. It is shown[28][29][30][31] that the adversary can take the fully advantage of the system configuration parameters to attack the system incurring large system mean square error.

Considering security, communication cost, and dynamic users; it is not feasible for the system to provide real-time response by sending all the received signals to the cloud. In most cases, it is infeasible to find optimal spectrum allocation strategies; however suboptimal solutions are fair enough to provide comparatively efficient system spectrum usage performance. As for the distributed cognitive radio network, each node has to coordinate to the nodes in the neighborhood in order to avoid the interference to the primary users. This will incur the communication burden and system security issue. Many cases need agile response in a real-time fashion to support the system functionality. For example[32], an agile and low cost atmospheric measurement system for energy harvest and real time mission support is developed such that a candidate sensor set can be dynamically deployed to wind filed estimation. In order balance the advantages and limitations between centralized and distributed frame, in this paper, we proposed *fog computing framework* in which fog nodes can analyzes the most time-sensitive data, close to the data sources and send the selected data or data summary for given time to the cloud for historical analysis and global optimization of the dynamic spectrum access.

## 2. SYSTEM MODEL

In this section, the system model is introduced to show how the DSA fog-computing framework can optimally detect and allocate the limited spectrum resources. Under this framework, each edge node can sense spectrum individually. Based on the rules provided by the cloud, each node has the ability to choose the best spectrum selection based on the reasoning process. Each node sends the data summary to cloud periodically for which the centralized cloud server can analyze the entire system and determine the updated rules for each edge node. Also, whenever there is abnormal data or a situation that the edge node cannot process, the data related to this certain situation is also sent to the cloud.

The DAS fog system structure can be seen in Figure 3, where each node can conduct its own spectrum sensing. As a sequence of signal is achieved, the fog node conducts feature extraction, anomaly detection, and decision making. The feature extraction includes waveform-based sensing, energy sensing, radio identification, cyclostationarity based sensing, and match filtering.

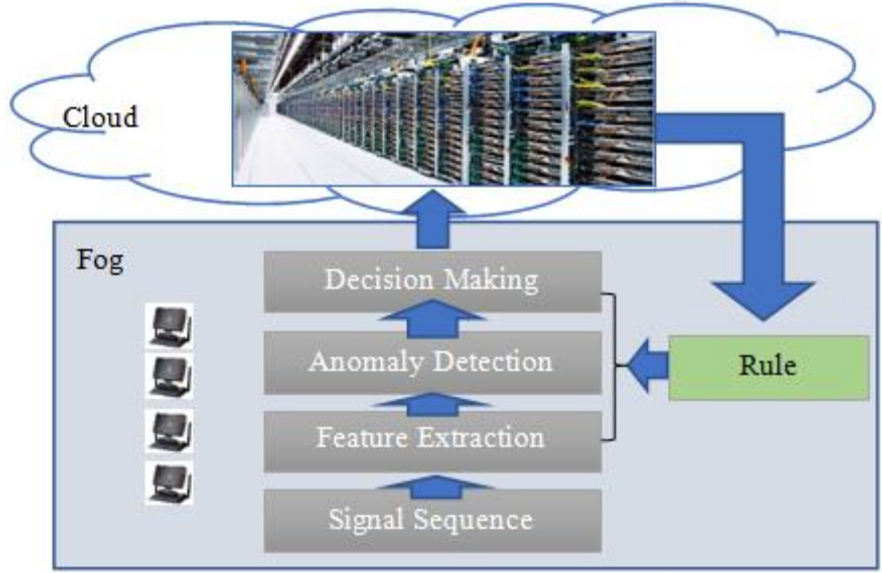

Figure 3 Fog Computing Framework

### 2.1 Machine Learning Approaches

All extracted features are used to determine the existence of spectrum and allocate the spectrum for the transmitting signals. Based on the computation capability of each fog node, different types of machine learning methods are incorporated in the framework such as least square regression, support vector machine (SVM), and manifold learning.

As for *least square logic regression*, it is supposed that vector $x$ denotes the training data, of which each element denotes the output from feature extraction process. It is supposed that $X = [x_1^T, x_2^T, ..., x_N^T]^T$ are the $N$ training samples. $y$ denotes whether certain spectrum is being used or not at the current time. The objective is to find the optimum $\hat{w}$ that can minimize the estimation error $||y - X\hat{w}||_2^2$. The objective function can be characterized as follows,

$$Min \; \frac{1}{N} ||y - Xw||_2^2 \quad (1)$$
$$st. ||w||_1 \leq k$$

where $k$ is the sparsity constraint. The corresponding Lagrangian formula can be rewritten as

$$Min \frac{1}{N} ||y - Xw||_2^2 + \lambda ||w||_1 \quad (2)$$

where $\lambda$ is penalty coefficient, $\lambda$ is independent of the $k$ in the primal formula. As $\lambda \to 0$, $\hat{w} = (X^T X)^{-1} X^T y$. In this case, the node which conducts the least square regression saves the parameter $w$.

As for the *Support Vector Machine (SVM)*,

$$\min_{w,b,\xi} \frac{1}{2} w^T w + C \sum_{i=1}^{n} \xi_i \quad (3)$$
$$st. y_i(w^T \phi(x_i) + b) \geq 1 - \xi_i,$$
$$\xi_i \geq 0, i = 1,2,...,n$$

The corresponding dual form is

$$\min_{\alpha} \frac{1}{2}\alpha^T H\alpha - \mathbf{1}^T\alpha \tag{4}$$

$$st. y^T\alpha = 0$$

$$0 \leq \alpha_i \leq C, i = 1, \ldots, N$$

where $\mathbf{1}$ is a column vector of all ones, $C$ is the upper bound, $H$ is an $N \times N$ semidefinite positive matrix, of which $H_{i,j} = y_i y_j K(x_i, x_j)$, the kernel $K(x_i, x_j) = \phi(x_i)^T \phi(x_j)$, kernel functions can be linear, polynomial, sigmoid, and Gaussian RBF, which can also be applied to the least square regression.

A third approach is *manifold learning*. When a non-linear relation exists between the features and desired output, each data $x_i$ is represented in a Euclidean space $x_i \in R^d$, where $d$ denotes the number of signal features. $x_i$ is assumed to be an image of some $\theta_n$ in a topological space. Let space $\Psi$ resemble $R^r$ with $r \leq d$. $\Psi$ is known as a manifold with reduction dimensionality $r$. Figure 4 illustrates the examples of 2-D manifolds embedded in a 3-D Euclidean space. The objective of manifold learning is to directly uncover the one-to-one map from $R^d$ to $R^r$. The mapping function should preserve the geometric structure of space $R^r$. Geometrically, this can be interpreted as uncurling a curved surface into a super-plane. Popular approaches for manifold learning include locally linear embedding[33], IsoMap[34], Laplacian eigenmap[35], and Maximum variance unfolding[36].

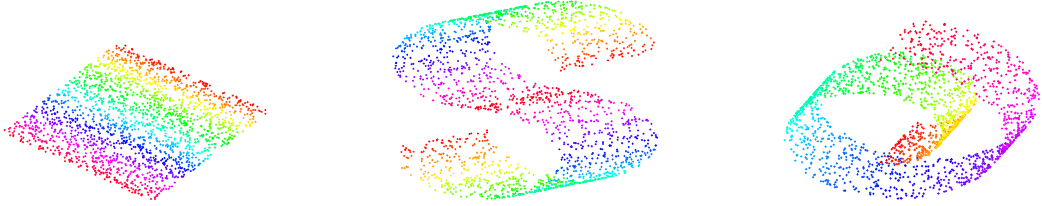

Figure 4 2-D Manifolds Embedded in A 3-D Space (Linear Subspace, S-curve, and Swiss roll)

Figure 5 shows the dimensionality reduction of the 3-D Swiss roll based on manifold learning. Figure 5 displays the results of several manifold functions, e.g., IsoMap, LLE, Hessian LLE, and provides the comparison results from linear approaches, such as Multidimensional scaling (MDS) and Principle component analysis (PCA). Since the 2D embedded manifold in Swill Roll is nonlinear, the linear approaches always fail to "uncurl" it. Also, in Figure 5, we can see that for manifold learning, IsoMap and Hessian LLE can successfully "uncurl" the original data, but LLE cannot. Hence, based on the computational capability of each edge node, manifold learning can be applied to feature extraction for dynamic spectrum access.

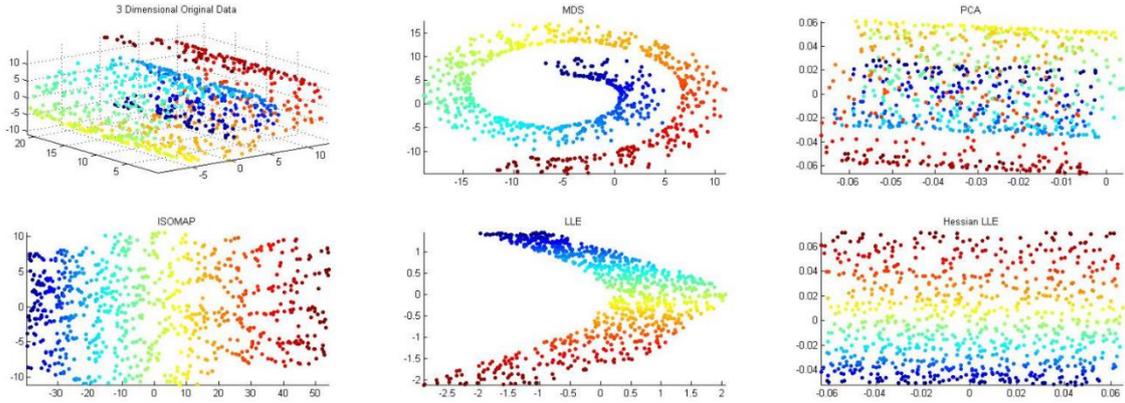

Figure 5 Dimensionality Reduction of 3-D Swiss Roll Based on Nonlinear Manifold Learning

# 3. SPECTRUM SENSING FOR FOG-BASED COGNITIVE RADIO

Spectrum sensing is a collective name of multiple techniques that are applied to cognitive radio to detect the unused frequency band and assign it to one or multiple secondary users. In this section, we briefly introduce and summarize some widely used signal processing techniques that have been reported in the literature and are suitable for our proposed fog framework, The methods determine whether the channel is occupied by the primary user and whether any unused spectrum is available for a secondary user. In the proposed framework, each fog node serves as a machine learning engine, and uses various outputs of a given spectrum sensing method based on the computational capability of each node.

## 3.1 Energy Detector

Energy detector-based sensing is the most widely used spectrum sensing algorithm due to its low computational complexity. Another advantage of energy assessment is that the energy detector is developed based on the assumption that receivers need no prior information of the primary user signals, which makes this sensing technique much more practical as compared to other existing approaches. However, in some cases, the second users have some information about the target to be detected. A popular Bayesian detector can be also utilized for detection[37].

The received signal is assumed as

$$y(n) = hx(n) + w(n) \qquad (5)$$

where $y(n)$ is the received signal, $h$ is the channel coefficient, $x(n)$ is the primary user's signal to be detected, $w(n)$ is the additive white Gaussian noise (AWGN), and $n$ is the sample index. Note that when $x(n) = 0$, there is no transmission from the primary user over this frequency band. The total energy observed over $N$ samples can be written as

$$M = \sum_{n=0}^{N} |y(n)|^2 \qquad (6)$$

Using (6), the spectrum occupancy decision can be obtained by comparing a threshold $\rho$, which is equivalent to the following hypothesis.

$$\begin{aligned} H_0: \quad & y(n) = w(n) \\ H_1: \quad & y(n) = hx(n) + w(n) \end{aligned} \qquad (7)$$

Equation (7) determines whether the primary user's signal occupies the spectrum. Note that $\rho$ is stored in "rule" block for fog processing, and $\rho$ can also be modified by the cloud in a real time fashion instead of a fixed value of a threshold used in the traditional energy detector sensing method.

## 3.2 Waveform-Based Sensing

In the case that the signal patterns are known to the receivers, the waveform-based sensing technique can be applied to the cognitive radio system. We use the same transmission model (5), from which the waveform-based sensing metric is written as

$$M = Re\left[\sum_{n=1}^{N} y(n)x^*(n)\right] \qquad (8)$$

where $*$ denotes the conjugate operation. When primary user signal presents on the spectrum, (8) can be further written as

$$M = h\sum_{n=1}^{N} |x(n)|^2 + Re\left[\sum_{n=1}^{N} w(n)\, x^*(n)\right] \qquad (9)$$

When primary user signal is absent, the sensing metric is written as

$$M = Re\left[\sum_{n=1}^{N} w(n)\, x^*(n)\right] \tag{10}$$

Similarly, a threshold value needs to be set in the system to make spectrum occupancy decision, and this value is modified by the cloud.

### 3.3 Cyclostationarity-Based Sensing

Cyclostationarity-based sensing is a technique that exploit the signal cyclostationarity features to detect the primary user transmissions. In this method, the cyclic correlation function is used to detect the signals in the frequency band. The cyclic spectral density function of a received signal is written as

$$S(f, \alpha) = \sum_{\tau=-\infty}^{\infty} R_y^\alpha(\tau) e^{-j2\pi f \tau} \tag{11}$$

where $\alpha$ is the cyclic frequency, and $R_y^\alpha$ is the cyclic autocorrelation function, which can be written as

$$R_y^\alpha(\tau) = E[y(n+\tau) y^*(n-\tau) e^{j2\pi \alpha n}] \tag{12}$$

When the cyclic frequency $\alpha$ equals the frequency of transmitted signal $x(n)$, the output of (11) reaches the peak value.

### 3.4 Feature Detection Sensing

Since primary user signal usually has its unique transmission pattern such as modulation scheme, carrier frequency, bandwidth, etc.; feature detection and machine learning algorithms can also be applied for spectrum sensing. This category of sensing techniques usually uses data-driven model rather than physical model to differentiate the primary user signal from the secondary user signal and noise. In this work, all types of features extracted by feature detection sensing are sent to the machine learning engine to make the decision on spectrum access.

## 4. CONCLUSION AND FUTURE WORK

In this paper, we propose a manifold learning dynamic spectrum allocation framework combining fog computing and cloud computing so that the received signal is processed close to where it is generated. Machine learning approaches are also incorporated to determine the available spectrum. Based on the features of the signal received, different machine learning methods are as applied such as least square logic regression, support vector machine, and manifold leaning. The feature extracted can also be projected to a higher feature space so that categorization performance can be further improved. Based on the rules defined, each fog node has the ability to reason and choose the best spectrum candidate to transmit the signal without interfering with the licensed legitimate primary users. The signal could be processed by the local nodes immediately, while the local node can also send the signal summary to the cloud so that the decision rules at the edge nodes rely on can be updated dynamically. Considering the computational capability of the cloud, the abnormal data from the edge nodes is also sent to the cloud for further analysis which enhances system security. The future work will include the hardware implementation regarding fog nodes of different computational capabilities and machine learning approach selection considering the feature of transmitted signal to optimize the dynamic spectrum allocation.